\documentclass[final,5p,times,twocolumn,authoryear]{elsarticle}
\usepackage{amssymb}
\usepackage{lipsum}
\usepackage{url}
\usepackage{amsmath}
\usepackage{booktabs}
\usepackage[colorlinks=true, allcolors=blue]{hyperref}
\usepackage{orcidlink}
\journal{High Energy Astrophysics}
\begin{document}
\begin{frontmatter}
%% Title, authors and addresses
%% use the tnoteref command within \title for footnotes;
%% use the tnotetext command for theassociated footnote;
%% use the fnref command within \author or \affiliation for footnotes;
%% use the fntext command for theassociated footnote;
%% use the corref command within \author for corresponding author footnotes;
%% use the cortext command for theassociated footnote;
%% use the ead command for the email address,
%% and the form \ead[url] for the home page:
%% \title{Title\tnoteref{label1}}
%% \tnotetext[label1]{}
%% \author{Name\corref{cor1}\fnref{label2}}
%% \ead{email address}
%% \ead[url]{home page}
%% \fntext[label2]{}
%% \cortext[cor1]{}
%% \affiliation{organization={},
%%            addressline={}, 
%%            city={},
%%            postcode={}, 
%%            state={},
%%            country={}}
%% \fntext[label3]{}

%% use optional labels to link authors explicitly to addresses:
%% \author[label1,label2]{}
%% \affiliation[label1]{organization={},
%%             addressline={},
%%             city={},
%%             postcode={},
%%             state={},
%%             country={}}
%%
%% \affiliation[label2]{organization={},
%%             addressline={},
%%             city={},
%%             postcode={},
%%             state={},
%%             country={}}

%\title{Study of gamma ray luminosity of pulsars under synchro-curvature regime}
\title{A synchro-curvature treatment of gamma-ray luminosity trends in pulsars}
\author[first,second]{A. Pathania\corref{cor1}\orcidlink{0009-0002-9979-9424}}
\ead{apathania@barc.gov.in}

\cortext[cor1]{Corresponding author}

\author[first,second]{K. K. Singh \orcidlink{0000-0002-5818-8195}}
\ead{kksastro@barc.gov.in}
\author[first,second]{K. K. Yadav}
\ead{kkyadav@barc.gov.in}

\affiliation[first]{organization={Astrophysical Sciences Division},%Department and Organization
            addressline={Bhabha Atomic Research Centre},
            city={Mumbai},
            postcode={400085},
            state={Maharashtra},
            country={India}}

\affiliation[second]{organization={Homi Bhabha National Institute},%Department and Organization
            addressline={Anushaktinagar},
            city={Mumbai},
            postcode={400094},
            state={Maharashtra},
            country={India}}

%\maketitle
\begin{abstract}
In recent years, the Large Area Telescope (LAT) onboard the \emph{Fermi} satellite has detected more than 300 pulsars in the high energy range. 
The population studies of high energy pulsars show that the gamma ray luminosity of a pulsar ($L_\gamma$) can be expressed in terms of the 
spin down luminosity ($\dot{E}$) as $L_\gamma \propto \dot {E}^s$ having exponent $s\sim 0.68$. This high energy emission, assumed to originate 
far from the stellar surface and near the light cylinder, is usually studied in either purely curvature or purely synchrotron regime. In this 
work, we adopt a synchro-curvature radiation framework to understand the origin of gamma ray emission from the pulsar and its implications at 
the population-level. By comparing the observed cutoff energies of the differential gamma-ray spectra with the theoretical synchro-curvature 
predictions and enforcing radiation reaction approximation, we determine the equilibrium Lorentz factor and pitch angle 
of the emitting charged particles. This approach allows to quantify the relative roles of curvature and synchrotron radiation to the radiative losses, 
thereby providing a physically grounded interpretation of the luminosity trend across the pulsar population.
\end{abstract}

%%Graphical abstract
%\begin{graphicalabstract}
%\includegraphics{grabs}
%\end{graphicalabstract}

%%Research highlights
% \begin{highlights}
% \item Research highlight 1
% \item Research highlight 2
% \end{highlights}

\begin{keyword}
%% keywords here, in the form: keyword \sep keyword, up to a maximum of 6 keywords
gamma rays \sep pulsars \sep radiation mechanism \sep synchro-curvature  \sep population studies \sep stars

%% PACS codes here, in the form: \PACS code \sep code

%% MSC codes here, in the form: \MSC code \sep code
%% or \MSC[2008] code \sep code (2000 is the default)

\end{keyword}

\end{frontmatter}

%\tableofcontents

%% \linenumbers

%% main text

\section{Introduction}
\label{introduction}
Pulsars are rapidly rotating strongly magnetized neutron stars which emit across the electromagnetic spectrum ranging from radio to gamma rays. 
These objects are normally detected by the radio telescopes, when the pulsed radio emission from these objects crosses the line of sight of the 
observer. Currently, more than 3800 pulsars have been tabulated in the Australian Telescope National Facility (ATNF) catalog \footnote{https://www.atnf.csiro.au/research/pulsar/psrcat/}\citep{manchester2005}, majorly populated in the galactic plane of the Milkyway galaxy \citep{pathania2023}. Out of the total pulsar population 
detected so far, about 10\% have been discovered in the high energy band \citep{smith2023} by the space-based Large Area Telescope (LAT) onboard the \emph{Fermi} 
satellite. The \emph{Fermi}-LAT, launched in 2008, has been continuously surveying the universe in the energy range from $30$ MeV to $> 300$ GeV \citep{atwood2009}. 
With its wide field of view of $2.4$ sr and better source localization and characterization than the earlier Energetic Gamma-Ray Experiment 
Telescope (EGRET) \citep{atwood2009,michelson2010}, LAT has detected the largest number of high energy pulsars. In recent past, the LAT collaboration published 
a number of pulsar catalogs (PCs) with 1PC consisting of 46 pulsars based on 6 months of data, and 2PC with 132 pulsars \citep{abdo2013}. The most recent up 
to date pulsar catalog (3PC) is based on 12 years of data and consists of a total of 340 well tabulated gamma-ray pulsars and candidates \citep{smith2023}.
\par
The rotation of a pulsar results in a very stable periodic pulsed signal. This periodic signal suffers a slow increase in its pulse period, except for 
occasional glitches \citep{shapiro1983}. Hence, a pulsar is characterized by its spin period $P$ and period derivative $\dot P$, both are measured very 
precisely using radio telescopes. Assuming the pulsar to be an orthogonal rotating magnetic dipole, one can have estimates of various quantities 
like surface magnetic field strength $B_s = \sqrt{\frac{3c^3I}{8\pi^2 R^6_{NS}}P\dot{P}}$, spin-down luminosity $\dot E = 4\pi^2 I \dot P / P^3$, 
characteristic age $\tau_c = P/2\dot P$ etc, where $I$ is the moment of inertia of the neutron star, and $R_{NS}$ is the radius of neutron star \citep{lorimer2004}. 
The observed global parameters in case of rotation powered pulsar decide the maximum extracted electromagnetic energy from the pulsar and structure of magnetosphere, 
which is populated by electron–positron plasma supplied through pair production cascades. \\
\indent Apart from efforts being made for detecting these faint gamma ray pulsars in the Galaxy, the current observational consensus suggests that radio and high-energy 
emissions originate from distinct regions of the magnetosphere. Radio emission is generally believed to be produced at relatively low altitudes above the polar 
caps through coherent plasma processes \citep{kramer1997,kijak1998}, whereas high-energy emission is attributed to incoherent radiation from ultra-relativistic 
particles accelerated in regions where an electric field parallel to the magnetic field develops\citep{vigano2015}. However, the precise location, 
geometry, and physical nature of these acceleration regions remain subjects of active debate. Various high energy emission models for pulsars, developed in the past, 
can be classified based on the assumed location of acceleration region. The Polar Cap model \citep{ruderman1975,harding1998} assumes the accelerating region 
near the stellar surface, whereas the same is extended to high altitudes in Slot gap model \citep{arons1983,muslimov2003}. Finally in the Outer-gap model 
\citep{cheng1986,romani1996,hirotani2008}, the acceleration region is located in the outer magnetosphere near the light cylinder. 
All these models, collectively falling under the magnetospheric scenario, are based on assumption of the existence of accelerating gaps. 
Electric field within these gaps develops a component along the magnetic field and results in the acceleration of charged particles. However, alternative to 
these models, stripped wind from the pulsar outside the light cylinder produces a thin Equatorial Current Sheet (ECS) layer and can also be a possible site for 
high energy emission via dissipative mechanism like magnetic reconnection\citep{lyubarskii1996,petri2012,mochol2017}.
\par
Recent results from the 3PC catalog indicate that the energy spectrum of pulsars follows a power law with sub-exponential cutoff \citep{smith2023} with 
cutoff energy lying at few GeV \citep{smith2023,pathania2023}. The sub-exponential nature of the energy spectrum and the observed pulse profile features 
(non-coincident radio and gamma profiles) \citep{romani1996,johnson2014}, favor the location of acceleration region in the outer magnetosphere, towards the 
light cylinder i.e. Outer gap model \citep{vigano2015}. The alternative to outer gap scenario, the existence of ECS in the wind stripped model at and beyond 
light cylinder, has also been verified from modeling of dissipation-less force-free magnetosphere of pulsars \citep{contopoulos1999,spitkovsky2006}. 
Particle-in-cell simulations and dissipative macroscopic solutions of pulsar magnetosphere also reveal high energy emission to take place near the ECS at 
light cylinder \citep{kalapotharakos2012,cerutti2016}.
\par
The high energy pulsar emission models differ in geometry and physical assumptions but share a common problem of connecting local particle acceleration and 
their radiative losses with the physically observables like gamma ray luminosities and cutoff energies. In short, the mode of radiative loss by the accelerated 
charged particles, either through curvature or synchrotron alone or combination of both, still remains unclear. Empirical studies of gamma-ray pulsar population 
and their correlations can help to connect the underlying physics with the emission processes. Empirical correlations between gamma ray luminosity, cutoff energy, 
and spin-down luminosity, including the so-called fundamental plane relations, indicate that pulsars occupy a restricted region of parameter 
space \citep{kalapotharakos2019}. These correlations encode information about the underlying acceleration mechanism and radiative regime, but their physical 
interpretations depend sensitively on the assumed radiation process and the treatment of radiation–reaction effects.
\par
In this work, we adopt a synchro-curvature radiation framework to understand the origin of pulsed gamma ray emission and its population-level implications under 
the radiation reaction approximation. First, we discuss the synchro-curvature formalism and corresponding gamma ray luminosity in 
Section \ref{synchrocurvature}. The dataset and corresponding methodology are discussed in section \ref{methodolgy}. Finally, the results and conclusions are 
presented in section \ref{results} and section \ref{conclusions} respectively.

\section{Synchro-curvature treatment of gamma-ray luminosity}\label{synchrocurvature}
\subsection{The synchro-curvature formalism}
The radiation emitted by relativistic charged particles moving along curved magnetic field lines with a finite pitch angle is described by 
the synchro-curvature formalism \citep{cheng1996,zhang1997,vigano2015b,torres2018}. The synchro-curvature framework unifies both curvature 
and synchrotron radiation and is applicable when both the guiding-center curvature and the gyrational motion around the magnetic field contribute to 
the particle acceleration \citep{vigano2015b}.
\par
We know that, when a charged particle (an electron or positron having electric charge $e$ and mass $m_e$) with Lorentz factor $\gamma$ moves in a 
magnetic field $B$ with pitch angle $\alpha$ relative to the local field direction, it will gyrate around the magentic field lines with the gyroradius 
$r_g$ where
\begin{equation}
r_g = \frac{\gamma \beta m_e c^2 \sin\alpha}{e B} ~~~~~~~~~~; ~~\gamma = (1-\beta^2)^{-1/2} ~~~\& ~~~\beta = v/c
\end{equation}
and $c$ is the speed of light  and $v$ is the velocity of the charge particle (taken to be c for ultra relativistic charged particle) . If the magnetic field line itself has radius of curvature $R_c$, this gyrational 
motion will slide along the curved magnetic field line where formar gives the synchrotron power loss whereas later gives the curvature energy loss. The relative 
importance of curvature and synchrotron effects is quantified by a dimensionless parameter called synchro-curvature parameter ($\zeta$) 
defined as \citep{vigano2015b} 
\begin{equation}
\zeta \equiv \frac{R_c}{r_g}\tan^2\alpha 
\end{equation}
It compares the curvature radius of the guiding center (magnetic field lines) to the effective transverse scale of the gyrational motion. Hence, the synchro-curvature 
regime can be expressed in terms of two functions of $\zeta$, as:
\begin{itemize}
\item $Q_2(\zeta)$, which modifies the characteristic photon energy,
\item $g_r(\zeta)$, which modifies the total radiated power.
\end{itemize}
These functions are defined as \citep{vigano2015b}
\begin{equation}
Q_2(\zeta) =
\frac{\cos^2\alpha}{R_c}
\left(1 + 3\zeta + \zeta^2 + \frac{r_g}{R_c}\right)^{1/2},
\end{equation}
and
\begin{equation}
g_r(\zeta) =
\left(\frac{R_c}{R_{eff}}\right)^2 \left(
\frac{1 + 7(R_{eff}Q_2)^{-2}}{8(R_{eff}Q_2)^{-1}}\right)
\end{equation}
where the effective curvature radius $R_{eff}$ is defined as
\begin{equation}
R_{eff} = \frac{R_c}{\cos^2\alpha \left(1 + \zeta + r_g/R_c\right)}
\end{equation}
The characteristic photon energy of synchro-curvature radiation is given by
\begin{equation}
\label{E_c}
E_c = \frac{3}{2}\hbar c\, \gamma^3 Q_2
\end{equation}
Under the synchro-curvature regime $(r_g \ll R_c)$, the characteristic photon energy formula reduces to the standard curvature radiation 
cutoff in the limit $\zeta \ll 1$ as,
\begin{equation}
E_{c,\rm curv} = \frac{3}{2}\hbar c\,\frac{\gamma^3}{R_c},
\end{equation}
and to standard synchrotron radiation cutoff in the limit $\zeta \gg 1$ as 
\begin{equation}
E_{c,\rm syn} = \frac{3}{2}\hbar c\frac{\gamma^3}{r_g} sin^2(\alpha) =  \frac{3}{2}\hbar \frac{\gamma^2 eB}{\beta m_e c} sin(\alpha)
\end{equation}
The total synchro-curvature radiative power loss of an electron is given as
\begin{equation}
P_{\rm sc} = \frac{2}{3} e^2 c\,\frac{\gamma^4}{R_c^2}~ g_r \equiv  P_{curv} ~g_r
\end{equation}
In the curvature-dominated regime ($\zeta \ll 1$), $g_r \rightarrow 1$ and the expression reduces to the standard curvature radiation power loss. 
While in the synchrotron-dominated regime ($\zeta \gg 1$), $g_r$ increases rapidly, indicating the radiative losses dominated by synchrotron-like 
transverse acceleration.
\par
However, the functional form of $g_r$ that can fit the exact formalism is given as \citep{vigano2015b}
\begin{equation}\label{gr} 
    g_r = 1 + 1.68\zeta + \zeta^2
\end{equation} 
and one can obtain limiting values of $\zeta$ for which the radiative power loss mechanism is said to be either curvature or synchrotron dominated. 
We define a given component to be dominant if radiative power loss by that mechanism constitutes $\sim 99\%$ of the total power loss. Hence, we have 
\begin{itemize}
\item Curvature dominated radiation: $P_{sc} \le 1.01 P_{curv} \Longrightarrow g_r \lesssim 1.01  \Longrightarrow \zeta \lesssim 0.006$
\item Synchrotron dominated radiation: $P_{sc} \ge 100 P_{curv} \Longrightarrow g_r \gtrsim 100 \Longrightarrow \zeta \gtrsim 9.1 \sim 10$
\end{itemize}
However, between the two asymptotic limits as discussed above, the intermediate regime characterized by $\zeta \sim 1$ is called the Synchro-curvature regime, 
where neither pure curvature nor pure synchrotron radiation alone provides an adequate description of the emission. In this case, the curvature of the magnetic field 
lines and the transverse gyrational motion contribute comparably to both the radiative losses and the spectral cutoff. Thus, we have
\begin{itemize}
\item Synchro-curvature regime: $1.01 P_{curv} \le P_{sc} \le 100 P_{curv} \Longrightarrow 1.01\lesssim g_r \lesssim 100  
	\Longrightarrow 0.006\lesssim \zeta \lesssim 10$

\end{itemize}
\subsection{Evaluating gamma ray luminosity}
The acceleration regions contain an electric field ($E_\parallel$) parallel to the magnetic field that accelerates the charged particle. Under the radiation reaction approximation, the charge particle emitting radiation will suffer an additional force called the radiation-reaction force which acts opposite to the emission direction. Here we have considered a specific scenario where an equilibrium between the radiation reaction force and the acceleration force existed such that, the energy gain of a charged particle from an electric field 
is balanced by its radiative loss, i.e. 
\begin{equation}\label{powerloss}
    eE_{||}c ~cos(\alpha) = \frac{2}{3}e^2 c \left(\frac{\gamma^4}{R_c^2}\right) g_r = P_{sc} 
\end{equation}
and same approximation is considered throughout this work. Equation (\ref{powerloss}) is used to evaluate the power loss of a charged particle that has been accelerated by an electric field $E_{||}$. 
A charged particle having Lorentz factor $\gamma$ radiates a photon spectrum that peaks close to the characteristic energy $E_c$. Since the particle emits synchro-curvature radiation, it will suffer a decrease in the pitch angle by losing the 
perpendicular momentum due to the synchro-curvature losses. However, theoretically, there are possibilities by which the particles can be supplied 
with a finite pitch angle by means of resonant absorption of radio photons \citep{lyubarskii1998,harding2008} and turbulent magnetic reconnection 
\citep{xu2022} etc. Hence, assuming the equilibrium condition within the radiation reaction approximation, the simultaneous solution to both equations (\ref{E_c}) and (\ref{powerloss}) 
helps to evaluate the equilibrium Lorentz factor and equilibrium pitch angle of the charged particle.
\par
The total gamma-ray luminosity of pulsar can be evaluated as product of number of accelerated charged particles ($N$) that participate in the high energy 
emission and radiative power loss by each particle ($P_{sc}$) as 
\begin{equation}\label{Lgamma}
L_\gamma = N P_{sc}
\end{equation} 
The number of accelerated charged particles can be parameterized in terms of particle flux originating along the open magnetic field lines. The effective rate of  radiating charged particles ($\dot{N}$) along the open magnetic field lines of aligned rotator can be estimated in terms of Goldreich-Julian 
density at the stellar surface ($n_{GJ}^{\star} = \frac{\Omega B_s }{2 \pi e c}$)\citep{goldreich1969}, area of polar cap ($A_{pc}$), and pair multiplicity 
($\kappa$), where $\kappa$ is initially assumed to be $1$. The area of polar cap can be estimated for an aligned rotator as $A_{pc} \sim \pi (R_{NS} \theta_{pc})^2$, 
where $\theta_{pc}$ is the polar cap angle defined as $\theta_{pc} \sim \sqrt{\frac{R_{NS}}{R_{LC}}}$, where $R_{LC} = cP/2\pi$ is the light cylinder radius. 
Therefore, we can write 
\begin{equation}
    \label{Ndot}
    \dot{N}= \kappa ~ n_{GJ}^\star~(\pi \frac{R_{NS}^3}{R_{LC}}) c
\end{equation}
 For acceleration region located at distance $r$ from the center of neutron star, the natural length scale turns out to be $r$. So, taking into account both, maximum physical size ($\sim r$) and radiation cooling effect, the number of particles that contribute to the gamma-ray luminosity is given as:
\begin{equation}
    \label{N}
    N = \dot{N} \times min(t_{ad},t_{cool})
\end{equation}
where $t_{ad} = r/c $ is the time for which the charged particle remains in the emission region and $t_{cool} = \gamma m_e c^2/P_{sc}$ is the time in 
which the particle losses its energy.

\section{Methodolgy}
\label{methodolgy}
\subsection{Assumptions}
\begin{itemize}
\item The outer gap model and ECS scenario suggest that the acceleration regions are located near the light cylinder. Therefore, we have assumed that the emission 
	region is located at light cylinder with physical size of $\sim R_{LC}$.
\item The magnetic field configuration of pulsar is assumed to be dipolar i.e. $ B \sim r^{-3}$. The magnetic field configuration of pulsar can have higher order 
	contribution \citep{gebino2025}, but at light cylinder distance $R_{LC}$, the dominating component of the magnetic field turns out to be dipolar with 
	magnetic field strength $B_{LC} = B_s \times \left( \frac{R_{NS}}{R_{LC}} \right)^3$,  where $B_{LC}$ is the dipolar magnetic field strength 
		at the light cylinder.
\item All the radiating particles are assumed to be mono-energetic under the equilibrium condition within the radiation reaction approximation.
\item The value of parallel electric field is parameterized in terms of the magnetic field with its upper limit constrained by the 
	local magnetic field \citep{kalapotharakos2019}. For the accelerating region near the light cylinder, we can write
\begin{equation}
    \label{eta}
    E_{||} = \eta B_{LC} ~~~~~; ~~\eta \le 1
\end{equation}
The variable $\eta$ is related to transfield thickness of the gap in the outer-gap model \citep{hirotani2013,vigano2015}, whereas in case of ECS model, 
it is related to the reconnection rate \citep{petri2012,cerutti2016} with typical value in the range $\sim 0.1 - 0.3$ and upper limit of 1.
\end{itemize}
\subsection{Data-Set}
We have used 3PC catalog extensively in the present study. The 3PC catalog provides location (source coordinates), measured quantites like spin period, spin down 
rate, fitted spectral parameters, and integral spectral energy flux above 100 MeV (i.e. $\mathrm{G_{100}}$) etc along with other derived quantities such as 
magnetic field at surface and light cylinder, and gamma ray luminosity  for each pulsar\footnote{https://fermi.gsfc.nasa.gov/ssc/data/access/lat/3rd\_PSR\_catalog/} 
\citep{smith2023}. For each pulsar in the sample, we use the phase-averaged spectral cutoff energy $E_c^{obs}$ defined as 
\begin{equation}
    \label{Ecobs}
    E_c^{obs} = \left(\frac{b^2}{d}\right)^{(1/b)} E_0
\end{equation}
where $E_0$ is the pivot energy, and the spectral shape parameters  $b$ and $d$ are referred to as exponential index and spectral curvature 
respectively \citep{smith2023,pathania2023}. The values of $E_c^{obs}$ are derived using equation (\ref{Ecobs}) with fitted exponential index values when available and fixed as 
$b=2/3$ when not available. Since the cutoff energy of the differential spectrum is similar to the characteristic photon energy of the underlying synchro-curvature emission 
process, we use $E_c^{\rm obs}$ as a proxy for the characteristic energy in the present study.
\par
\begin{figure}[htbp]
    \centering
    \includegraphics[width=\linewidth]{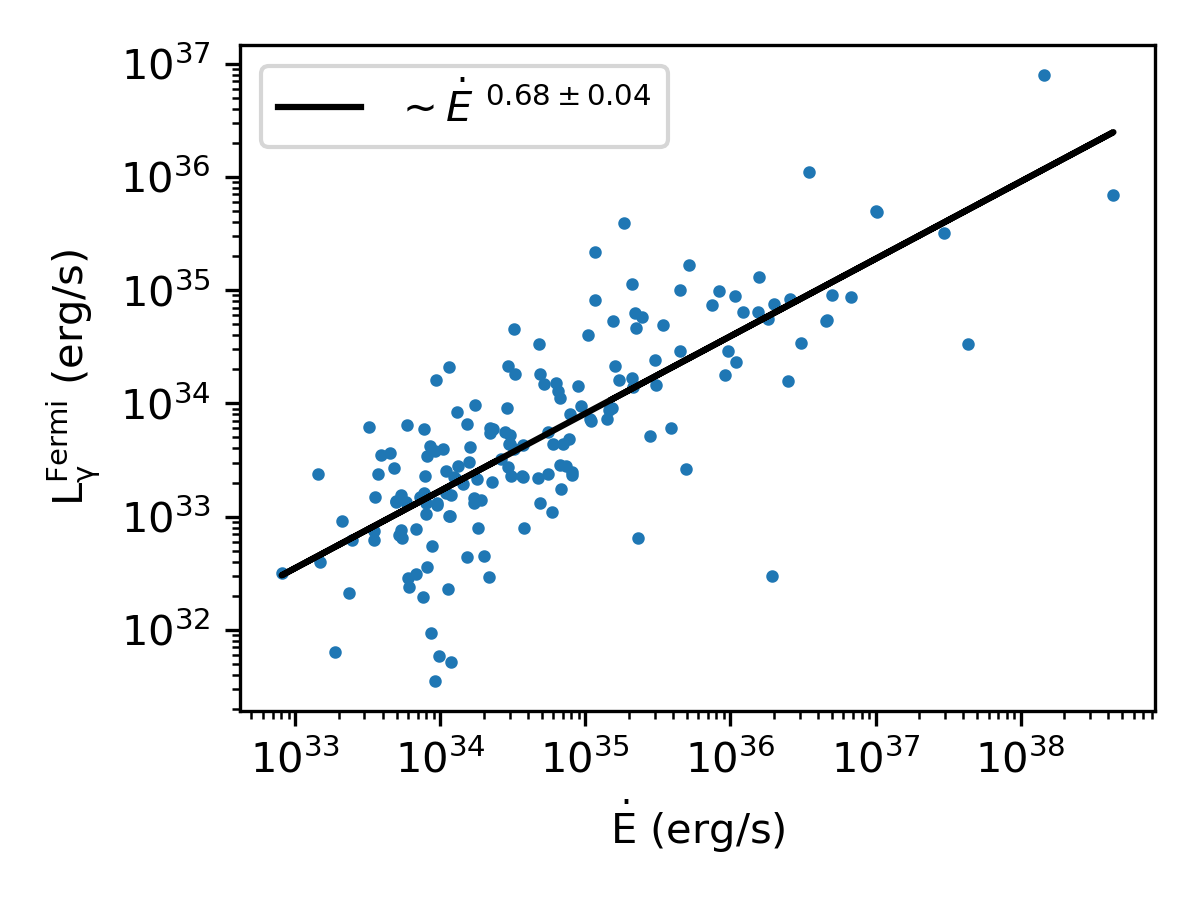}
\caption{Gamma ray luminosity of selected pulsars evaluated in energy band 0.1-100 GeV vs spin down luminosity $\dot E$. The solid black line shows the power law fit to the 
	data.}
\label{fig:fermi}
\end{figure}

The gamma ray luminosity of these pulsars is calculated as 
\begin{equation}
    \label{LUMG-formula}
    L_\gamma^{Fermi} = 4\pi d^2f_\Omega G_{100}
\end{equation}
where $f_\Omega$ is the beaming fraction and $d$ is the distance of the pulsar from the Earth. We have used $f_\Omega = 1 $ in this study for all pulsars.
Although, the distance to all pulsars are not known, we have selected only those pulsars for which distance and luminosity evaluations are 
given in the catalog and their corresponding cutoff energy lies in between 20 MeV to 100 GeV, results in total of 167 selected pulsars. The selected observed high energy pulsar population shows that the gamma-ray luminosity ($L_\gamma^{Fermi}$) has a correlation with spin down 
luminosity ($\mathrm{\dot{E}}$)  as $ L_\gamma^{Fermi} \propto \dot{E}^{0.68 \pm 0.04}$ with corresponding scatter of $0.57~ dex$ as shown in 
figure \ref{fig:fermi}.

\subsection{Evaluating pitch angle and Lorentz factor using Nelder-Mead algorithm}
Evaluation of equlibrium pitch angle and Lorentz factor of charged particles in the accelerating regions requires solution to both equations (\ref{E_c}) and (\ref{powerloss}). These two equation are non-linear in nature and roots of these equations can 
be obtained by making use of Nelder-Mead algorithm \citep{nelder1965}. Nelder-Mead algorithm is based on the geometrical figure of simplex 
and involves only evaluation of function values and not its derivatives. For a given problem with $X$ dimension (which are $\alpha$ and $\gamma$ in our case), 
it constructs $X$ dimensions of a triangle having $X+1$ vertices. Assuming the spectral cutoff energy and the characteristic photon energy to be comparable, 
the cost function of Nelder-Mead algorithm defined as 
\begin{equation}\label{cost_function}
C(\gamma,\alpha) = \left|\frac{E_c(\gamma,\alpha)-E_c^{\rm obs}}{E_c^{\rm obs}}\right|
+
\left|\frac{P_{\rm sc}(\gamma,\alpha)/cos(\alpha)-eE_\parallel c}{eE_\parallel c}\right|
\end{equation}
is evaluated at each vertex, located at a ($\gamma,\alpha$) pair from the parameter space and a motion in parameter space is promoted via simplex transformations such 
that cost function is minimized to less than $0.001$. Using this procedure, the equilibrium Lorentz factor and pitch angle for a given pulsar has been evaluated. The same 
procedure is being carried out for each pulsar to evaluate the corresponding $\gamma$ and $\alpha$ of the particle population subject to acceleration by electric field in the 
accelerating region. The Nelder-Mead algorithm is initially run on a coarse grid of $\gamma$ ranging from $5\times10^5$ to $5\times 10^7$ and $\alpha$ in the range 
$10^{-5}$ to $1$ radians subject to convergence below $0.1\%$.

\section{Results and Discussion}\label{results}
We discuss the results obtained by applying the synchro-curvature formalism under equilibrium condition within the radiation reaction approximation on the selected sample of 167 pulsars 
for which distance and luminosity estimates are available in the 3PC catalog. We have evaluated the equilibrium Lorentz factor ($\gamma$) and pitch angle ($\alpha$) by 
simultaneously satisfying equation (\ref{E_c}) and equation (\ref{powerloss}) using Nelder-Mead 
algorithm\footnote{https://docs.scipy.org/doc/scipy/reference/optimize.minimize-neldermead.html} for three different values of $\eta = (0.1, 0.3, 1.0)$ subject 
to convergence below $0.1\%$ and corresponding population level luminosity trends are also being explored. Out of 167 pulsars, the desired convergence has been 
achieved for $(114,144,163)$ pulsars for $\eta = (0.1, 0.3, 1.0)$ respectively. The representative contour plots of $(\gamma,\alpha)$ for few pulsars 
evaluated for different values of $\eta$ are shown in \ref{AppendixA}.

\subsection{Joint dependence of $\alpha$ and $\gamma$ on the accelerating field}
\begin{figure}[htbp]
    \centering
    \includegraphics[width=\linewidth]{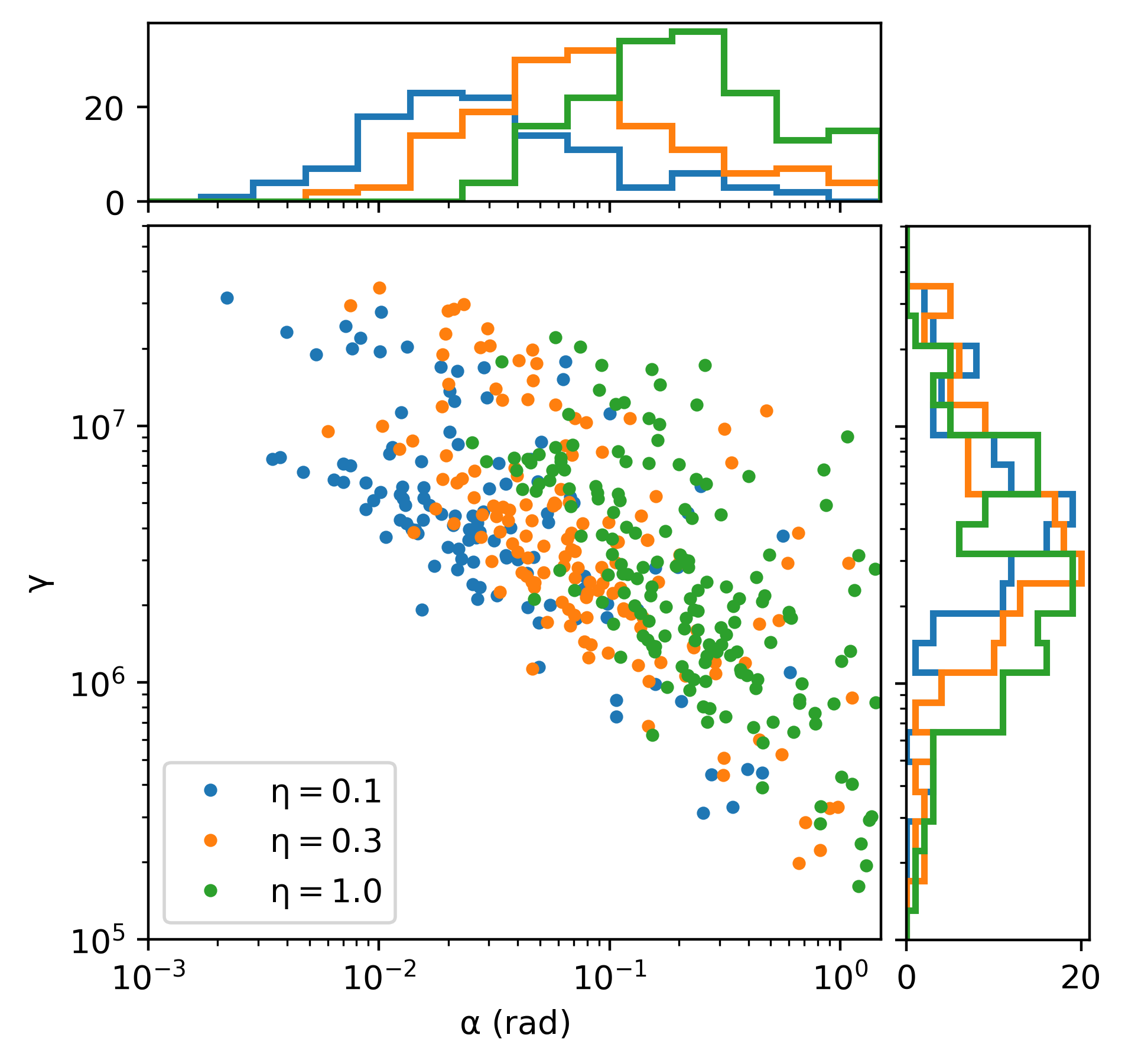}
\caption{Lorentz factor vs pitch angle, evaluated for three different values of electric field $E_{||} = \eta B_{LC}$, parametrized by parameter $\eta$. 
	The histograms of pitch angle and Lorentz factor for three different values of $\eta$ are shown on top and right respectively. The typical error in the estimation of pitch angle and Lorentz factor is less than 0.5\% of their absolute values.}
\label{fig:gammavsalpha}
\end{figure}
We have investigated the relationship between the evaluated equilibrium pitch angle ($\alpha$) and Lorentz factor ($\gamma$) for different values of 
the electric field present in the accelerating regions, parameterized by parameter $\eta$. For all explored cases, a clear anti-correlation is visible 
between $\alpha$ and $\gamma$, with higher Lorentz factors corresponding to smaller pitch angles and vice-versa as shown in figure \ref{fig:gammavsalpha}. This anti-correlation clearly suggests a significant contribution of synchrotron component in the total emission mechanism. It is observed that with increase in the value of $\eta$, which corresponds to increase in existed value of electric field in the accelerating regions, the equilibrium 
solutions shift toward relatively lower Lorentz factors and larger pitch angles and vice-versa. This behavior reflects the balance between acceleration and 
radiative losses. Since, the dynamics of charged particle in the presence of electric field is not solved explicitly, hence the presence of different $E_{||}$ within the acceleration region parametrize the different acceleration environments. When $E_\parallel$ is weaker, particles must reach higher Lorentz factors to reproduce the observed cutoff energies, which in turn enforces stronger 
pitch-angle damping. For stronger electric fields, acceleration is more efficient and the same cutoff energy can be achieved at lower $\gamma$, allowing larger pitch 
angles to persist. These results highlight the strong coupling between local acceleration conditions and particle pitch angle.

\subsection{Pitch angle vs cutoff energy}
\begin{figure}[htbp]
    \centering
    \includegraphics[width=\linewidth]{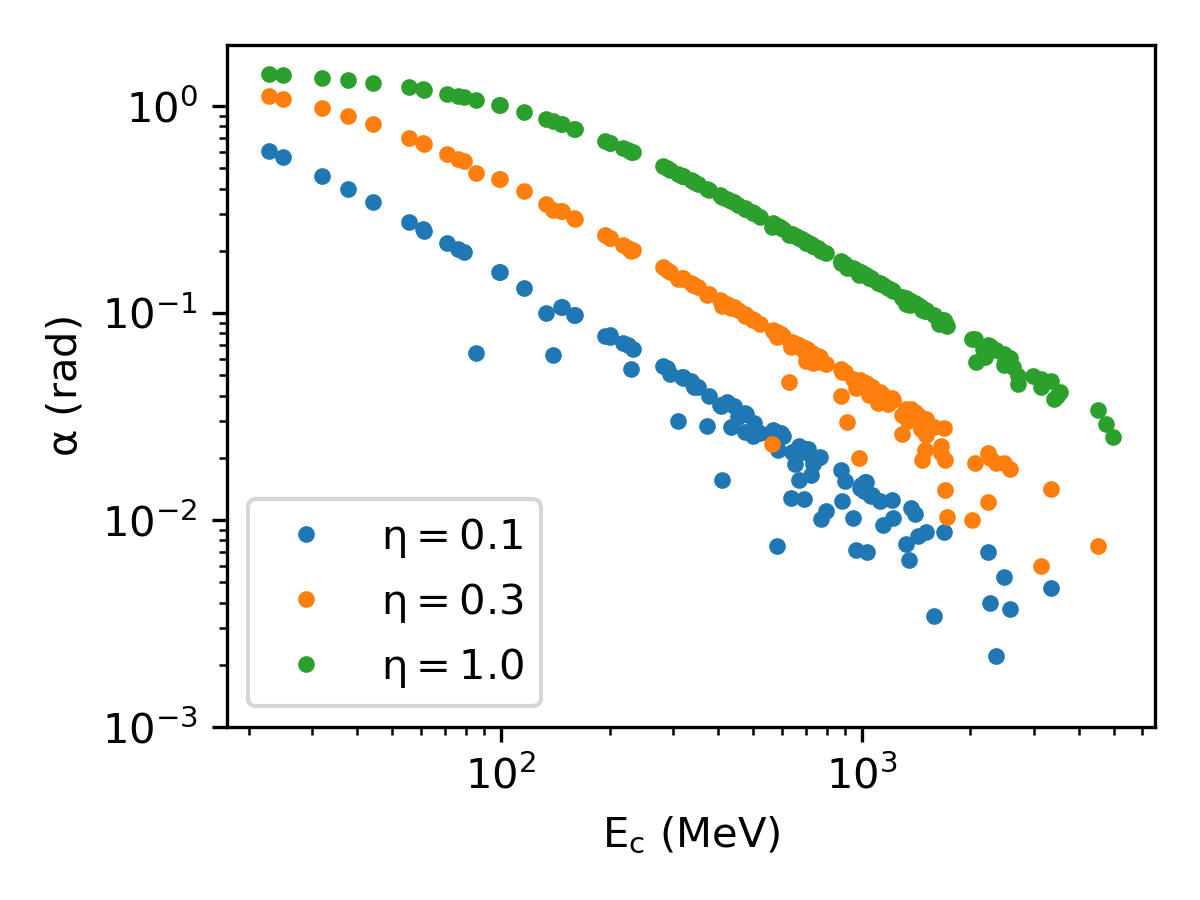}
\caption{Pitch angle vs observed cutoff energies for three different values of $\eta$.}
\label{fig:alpha}
\end{figure}

Examining the pitch angle ($\alpha$) with respect to cutoff energy ($E_c$) shows that the pitch angle decreases monotonically from $\alpha \sim 1$ rad to 
$\alpha \sim 10^{-3}$ rad as the cutoff energy increases across the pulsar sample as shown in figure \ref{fig:alpha}.
For a fixed value of accelerating field parametrized by $\eta$, higher cutoff energies require larger Lorentz factors 
and a more curvature-dominated trajectory. As $\gamma$ increases, even modest pitch angles lead to strong synchrotron losses, which rapidly decays in order to maintain the equilibrium condition within the radiation reaction approximation. Consequently, particles emitting at higher cutoff energies are driven toward smaller pitch angles, whereas lower 
cutoff energies permit larger equilibrium pitch angles. This result provides a physically grounded explanation for the relatively narrow 
range of cutoff energies observed in  gamma ray pulsars despite a wide spread in spin-down power, and is consistent with the notion that 
pitch-angle damping plays a central role in regulating the emission.

For a given cutoff energy, the presence of higher accelerating field strength in the acceleration region requires an equilibrium pitch angle 
to be relatively large under the equilibrium condition within the radiation reaction approximation. These higher pitch angles of charged particles will result in increase of synchrotron-like losses, 
allowing the particles to balance the larger acceleration power without increasing the cutoff energy. As a result, for higher values of $\eta$ present in the 
acceleration regions, the radiation mechanism shifts from synchro-curvature dominated region to the synchrotron dominated region.

\subsection{Synchro-curvature regime indicator $\zeta$}
\begin{figure*}[htbp]
    \centering
    \includegraphics[width=\linewidth]{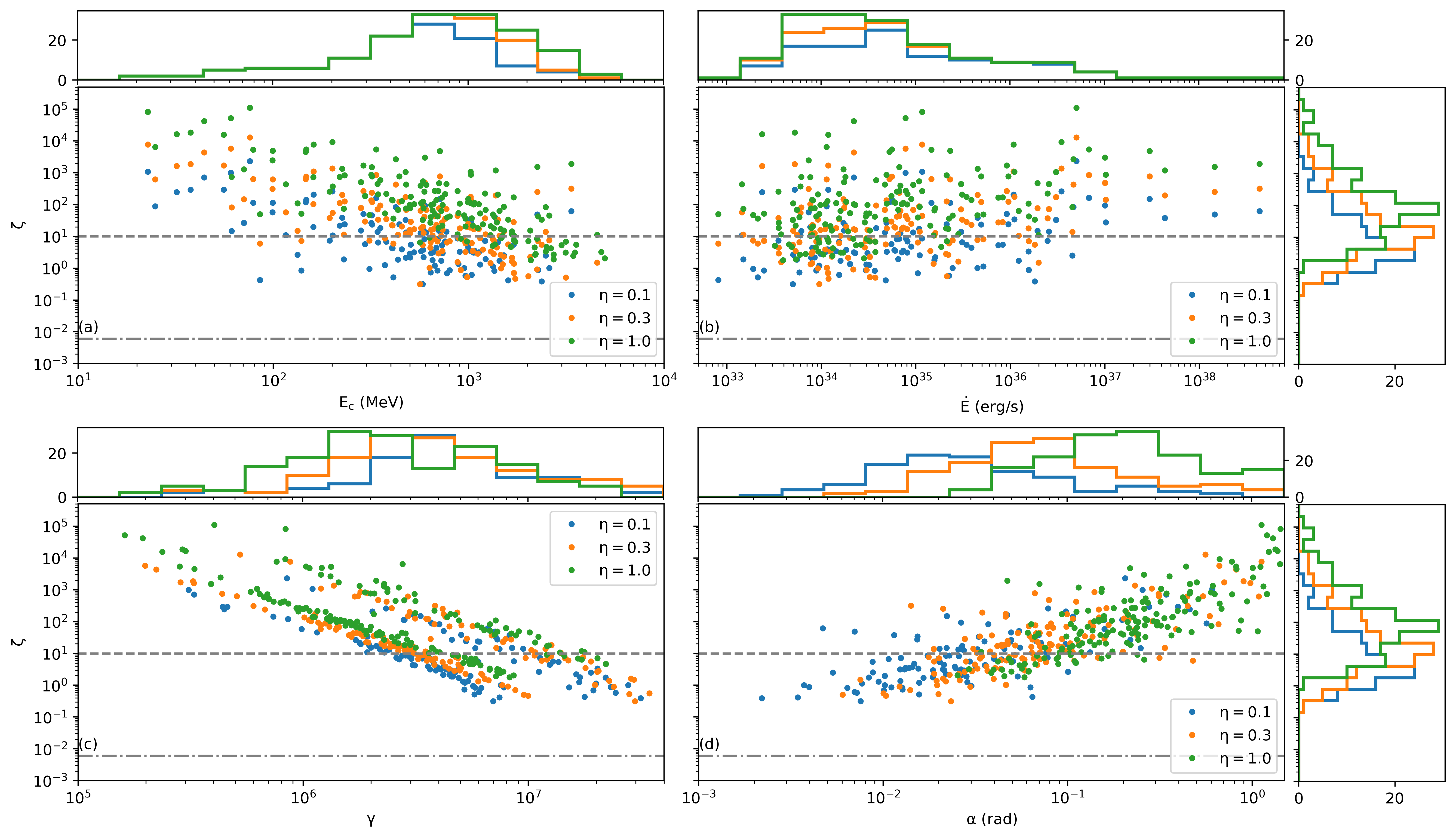}
\caption{The synchro-curvature parameter ($\zeta$) vs observed cutoff energy of pulsar ($E_c$), spin down luminosity of pulsar ($\dot E$), 
	Lorentz factor of particle ($\gamma$) and pitch angle of particle ($\alpha$) for three different values of $\eta$. The horizontal dashed and 
	dashed-dotted lines correspond to $\zeta = 10$ and $\zeta = 0.006$ respectively.}
\label{fig:zeta}
\end{figure*}

The relative importance of curvature and synchrotron components in the total radiative power loss can be studied with the help of the synchro-curvature parameter $\zeta$. 
We have plotted parameter $\zeta$ as a function of $E_c$, $\dot E$, $\gamma$, and $\alpha$ in figure \ref{fig:zeta}(a,b,c, and d) respectively. 
From figure \ref{fig:zeta}(a), we have found that $\zeta$ is typically large ($\zeta \gtrsim 10$) for pulsars with lower cutoff energies, indicating that 
their instantaneous radiative losses are dominated by synchrotron-like gyrational motion. In contrast, pulsars with higher cutoff energies tend to cluster 
around $\zeta \sim 0.5$-$1$, corresponding to the synchro-curvature regime in which both the curvature and synchrotron radiation effects become important. 
However, for a given cutoff energy, with increased value of $\eta$ or available accelerating electric field, the relative synchrotron component within the 
total radiative loss increases. A similar trend is observed with spin-down luminosity shown in figure \ref{fig:zeta}(b). Pulsars with lower $\dot E$ preferentially 
occupy the low-$\zeta$ regime, whereas higher-$\dot E$ pulsars exhibit systematically larger $\zeta$ values.  Moreover, from figure \ref{fig:zeta}(c,d), we find that 
the obtained equilibrium Lorentz factor of particles reduces and the equilibrium pitch angle increases as the radiation regime transits from less synchrotron dominated regime 
to more synchrotron dominated regime. The lower and upper cluster in figure \ref{fig:zeta}(c) correspond to the populations of millisecond ($P < 20$ ms) and young ($P \ge 20 $ ms) pulsars respectively and can be attributed to relatively smaller radius of curvature of magnetic field lines of millisecond pulsar population at the light cylinder with respect to that of normal pulsar population.

\subsection{Gamma-ray luminosity and inferred effective pair multiplicity}
\begin{figure}[htbp]
    \centering
    \includegraphics[width=\linewidth]{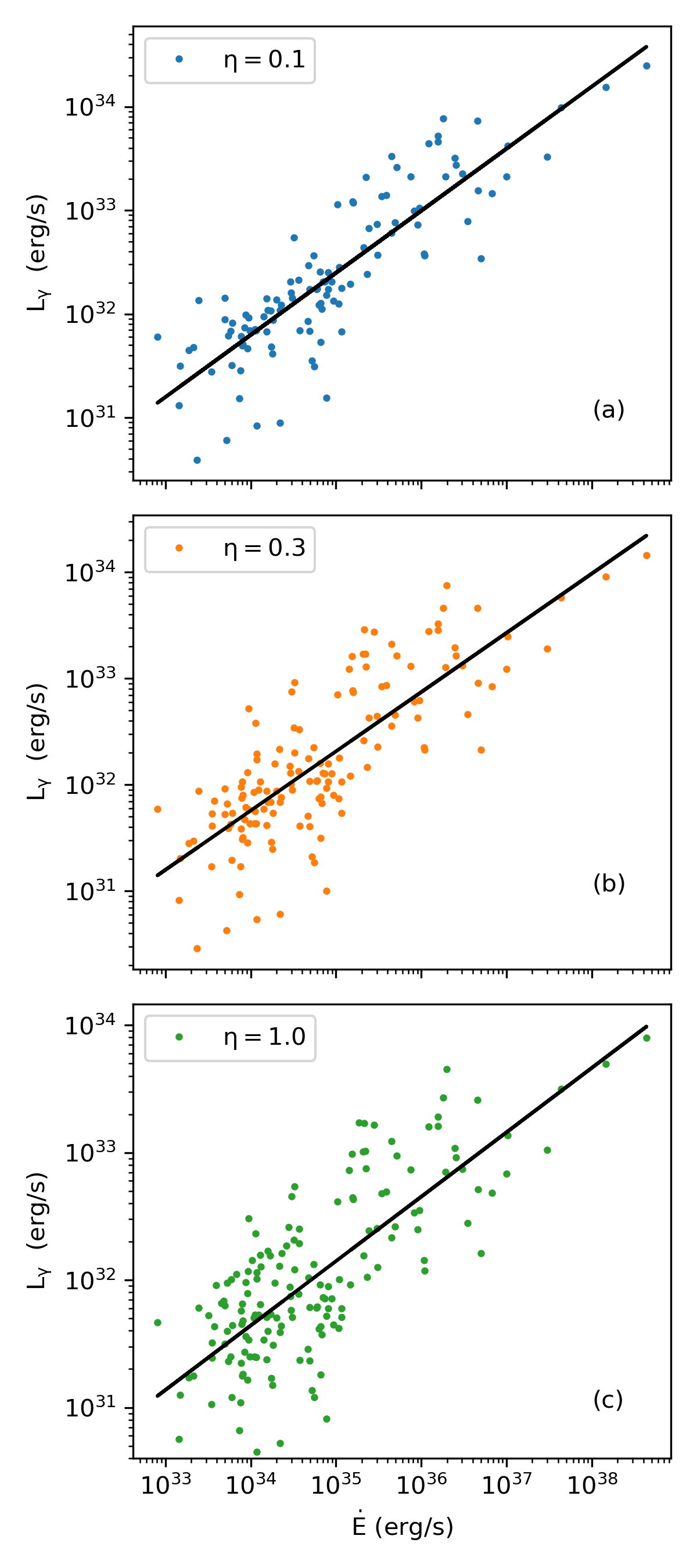}
\caption{Estimated gamma ray luminosity under synchro-curvature formalism as a function of spin down luminosity for three different values of $\eta$. 
	The solid line shows the fitted power law $L_\gamma \propto \dot E^s$ to the scatter.}
\label{fig:luminosity}
\end{figure}
Using the equilibrium Lorentz factor and pitch angle obtained from simultaneous solution to both  equation (\ref{E_c}) and equation (\ref{powerloss}), we compute the 
expected gamma ray luminosity using equation (\ref{Lgamma}) assuming a unit pair multiplicity ($\kappa = 1$). The resulting model luminosities as a function of spin down luminosity are shown in figure \ref{fig:luminosity}. For all explored values of $\eta$ as shown in figure \ref{fig:luminosity}, the 
model luminosities follow a power-law dependence with $\dot E$ (i.e. $L_\gamma \propto \dot{E}^{s\pm \Delta s}$) with exponent $s\sim 0.55$, consistent with 
the population-level trends derived from recent observations (figure \ref{fig:fermi}). The values of exponent and scatter around the best-fit relation for 
different $\eta$ are given in Table \ref{tab:exponentandkappa}. The scatter around the best-fit relation is approximately $0.40$ dex in all cases, indicating 
that the overall dispersion is largely insensitive to the choice of $\eta$ within the explored range. This implies that the observed luminosity scaling can be 
reproduced without invoking fine-tuned acceleration efficiencies whereas the persistence of comparable scatter across different $\eta$ values suggests that 
variations in the accelerating field alone cannot account for the full spread in observed luminosities.
\par
\begin{figure}[htbp]
    \centering
    \includegraphics[width=\linewidth]{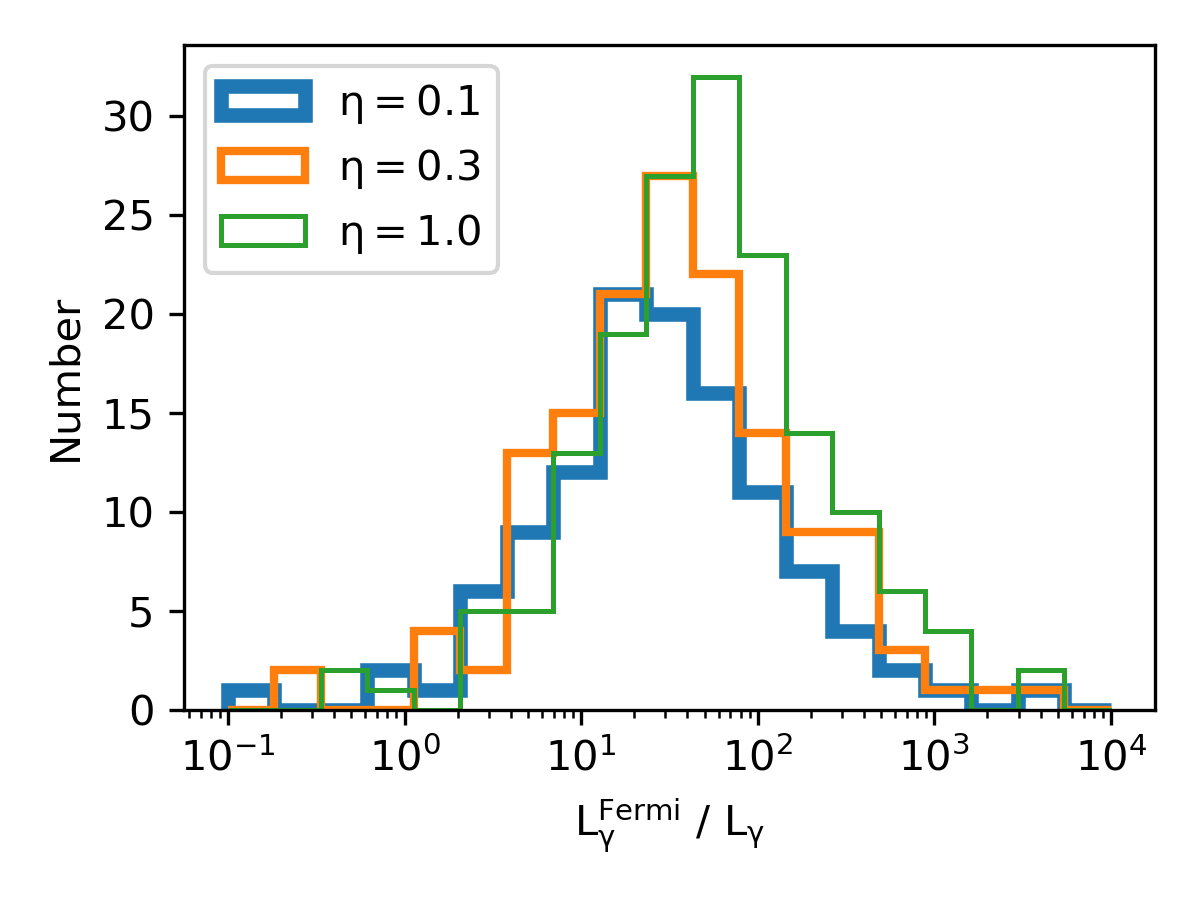}
\caption{Distribution of $L_\gamma^{Fermi} / L_\gamma$ for three different values of $\eta$.}
\label{fig:kappa}
\end{figure}

We also explore the effective pair multiplicity $\kappa$ by comparing the observed luminosity with the model luminosity obtained for $\kappa = 1$ given as 
\begin{equation}\label{eq:kappa}
\kappa = \frac{L_\gamma^{\rm Fermi}} {[N P_{sc} ]_{\kappa=1}}.
\end{equation}
The resulting multiplicities, spanning over a broad range across the pulsar population, can be attributed to the differences in pair production efficiency and 
plasma supply as shown in figure \ref{fig:kappa}. For a given value of $\eta$, one can plot the distribution of ratio of the observed luminosity with respect to the model luminosity and maximum of that distribution will give the representative value of $\kappa$.
This approach allows to estimate effective $\kappa$ without assuming a specific gap geometry or cascade model. The inferred multiplicity thus 
represents an effective, population-averaged quantity that accounts for both primary particle injection and secondary pair production within the emitting 
region that participate in the gamma-ray emission. The inferred $\kappa$ values as reported in Table \ref{tab:exponentandkappa} encapsulate the combined 
effects of magnetospheric conditions, pair cascade development, and particle injection rates, and therefore is termed as an effective pair multiplicity.
\begin{table}[!ht]
\centering
\begin{tabular}{|ccccc|}%{|p{2cm}p{3cm}p{3.5cm}p{3cm}p{3cm}|} 
\hline
Sr. No. & $\eta$ & s $\pm \Delta$s &  scatter (dex) & $\kappa$\\
\hline
1  &  0.1  & 0.60 $\pm$ 0.03  & 0.39 & 17 \\
2  &  0.3  & 0.56 $\pm$ 0.03  & 0.45 & 32 \\
3  &  1.0  & 0.50 $\pm$ 0.03  & 0.45 & 58 \\
\hline
\end{tabular}
\caption{Values of  fitted power law exponent  and scatter to the estimated gamma ray luminosity vs spin down luminosity and the inferred values of the effective pair multiplicity for three different values of $\eta$. }
\label{tab:exponentandkappa}
\end{table}

\section{Conclusions}\label{conclusions}
In this work, we have investigated the origin of gamma ray emission from pulsar magnetospheres within the synchro-curvature radiation framework. 
By combining the observed differential spectral cutoff energies with the equilibrium condition within the radiation reaction approximation, we solved for the equilibrium Lorentz 
factor and pitch angle of emitting particles and quantified the relative roles of curvature and synchrotron contributions in the total radiative losses.
The main results of this study are summarized below:
\begin{itemize}

\item We have used derivative-free Nelder-Mead optimization scheme to determine equilibrium solutions across the allowed parameter space with accuracy of less than $0.1\%$. The convergence properties of the solutions support the interpretation of the radiation-reaction equilibrium as a stable attractor in $(\gamma,\alpha)$ space.

\item Under the equilibrium condition within the radiation reaction approximation, equilibrium values of Lorentz factor and pitch angle have been derived under the synchro-curvature formalism for different values of accelerating electric field rather than assuming extreme limits of radiative loss i.e. either curvature or synchrotron radiation loss.

\item We found that for a wide range of pulsar parameters, the synchro-curvature parameter $\zeta$ typically exceeds unity, indicating that instantaneous radiative 
	losses are often dominated by synchrotron-like gyrational motion. However, higher cutoff energy pulsars with $\zeta \sim 1 $ imply that the corresponding radiation mechanism is purely synchro-curvature in nature.

%\item We showed that assuming $E_\parallel = \eta B_{\rm LC}$ with $\eta \lesssim 0.3$ yields physically consistent solutions compatible, while $\eta \sim 1$ should be regarded as a local upper bound applicable only to compact, charge--starved or strong reconnecting regions.

\item With increase in value of $\eta$ under the equilibrium condition within the radiation reaction approximation, constrained by the cutoff energy equation, the radiation mechanism shifts from synchro-curvature regime ($\zeta \sim 1$) towards the synchrotron dominated regime ($\zeta \sim 100$).

\item  For all explored values of $\eta$, the model luminosities follow a power-law dependence on $\dot E$ with exponent $s\sim 0.55$, with scatter around $\sim0.40$ dex, 
	reproducing global trends of observed gamma ray luminosity as a function of spin down luminosity. The exponent seems to be independent of accelerating 
		electric field within the explored range of $\eta$, whereas scatter can be attributed to various unknown quantities like beaming fraction, 
	uncertainties in pulsar distance estimates etc. \citep{pascual2025}.
\item We have estimated typical value of effective pair multiplicity $\kappa$ without assuming a specific gap geometry or cascade model. The value of $\kappa$ is 
	found to increase with increase in assumed accelerating electric field with a typical value lying in the range $\sim 20 - 60$.
\end{itemize}
Although we have evaluated the equilibrium Lorentz factor and pitch angle for the current pulsar population using the \emph{Fermi}-LAT 3PC catalog, the size of the 
gamma ray pulsar sample remains relatively low. The statistical robustness of population-level trends will improve with increase in the number of detected gamma ray 
pulsar population in the coming years like the release of $4^{th}$ pulsar catalog (4PC) by the \emph{Fermi}-LAT based on multi-wavelength follow ups and machine 
learning based hints \citep{zhu2024,pathania2026}. The expanded sample along with the combination of other waveband data will provide an opportunity to tightly 
constrain pair multiplicities, acceleration efficiencies, and magnetospheric dissipation processes. This will advance our understanding of the particle acceleration 
and high-energy emission in pulsar magnetospheres.
\section*{Acknowledgements}
Authors sincerely thank the anonymous reviewer for his/her thorough and insightful feedback on the manuscript. A. Pathania also thanks Mr. Gunindra Krishna Mahanta or academic discussions and suggestions. We would like to thank the Fermi Science Support 
Center (FSSC) for the public availability of data.

%% The Appendices part is started with the command \appendix;
\appendix
\section{Representative Contour plots for few pulsars}
\label{AppendixA}
\begin{figure*}[htbp]
    \centering
    \includegraphics[width=\linewidth]{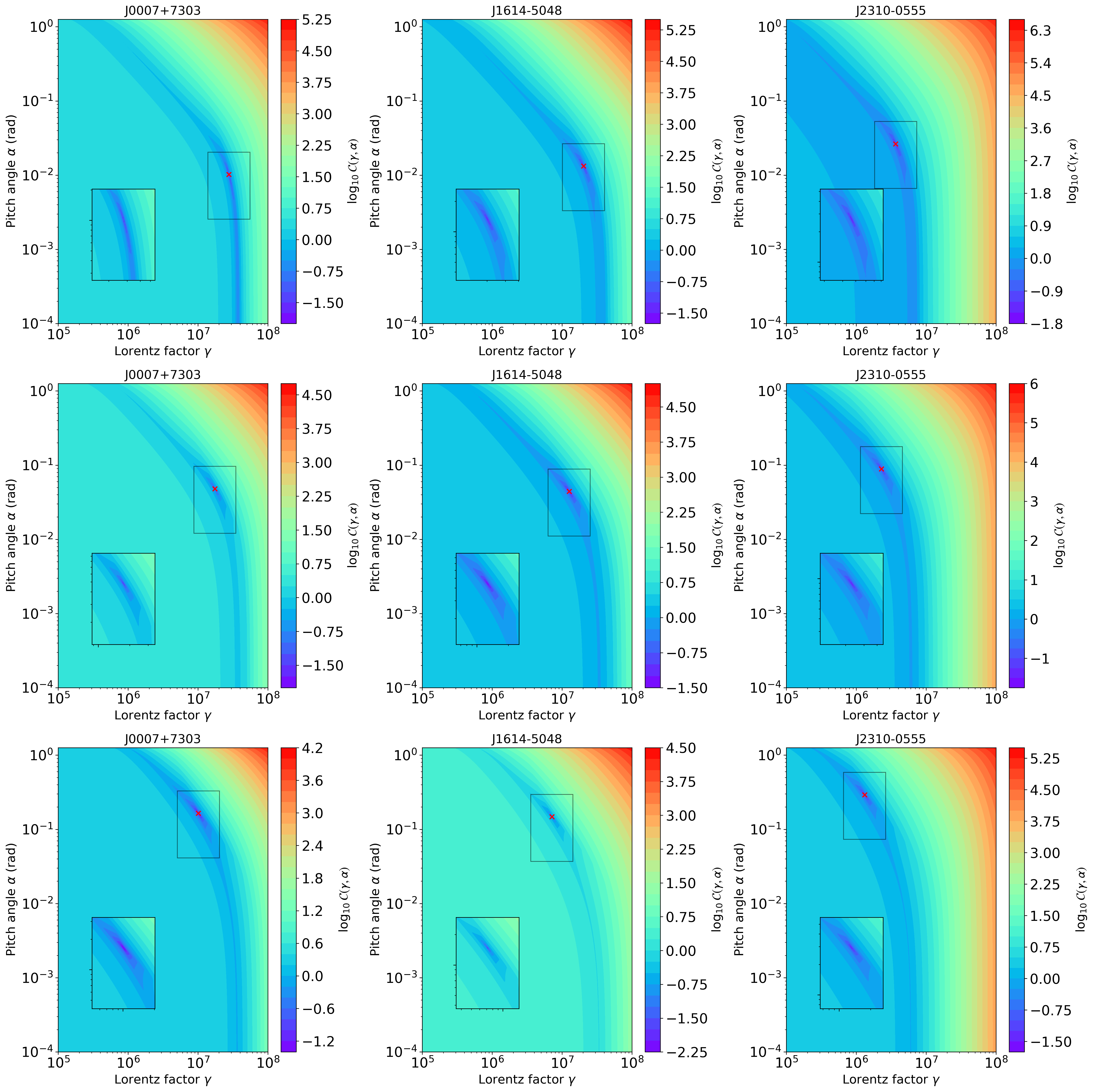}
\caption{Contour plots of $C(\gamma,\alpha)$ for 3 pulsars evaluated at different values of $\eta$. The top, middle and bottom row correspond to $\eta$ values of 0.1, 0.3 and 1.0 respectively. The red cross shows the equilibrium values of Lorentz factor and pitch angle of particles obtained by simultaneously satisfying the equations (\ref{E_c}) and (\ref{powerloss}) using Nelder-Mead algorithm whereas the rectangle on each contour plot is a zoom around the minimum. }
\label{fig:Costfunction}
\end{figure*}
%% appendix sections are then done as normal sections

%% If you have bibdatabase file and want bibtex to generate the
%% bibitems, please use
%%
\bibliographystyle{elsarticle-harv} 
\bibliography{biblography}

%% else use the following coding to input the bibitems directly in the
%% TeX file.

%%\begin{thebibliography}{00}

%% \bibitem[Author(year)]{label}
%% For example:

%% \bibitem[Aladro et al.(2015)]{Aladro15} Aladro, R., Martín, S., Riquelme, D., et al. 2015, \aas, 579, A101

%%\end{thebibliography}

\end{document}